\documentclass[aps,floatfix,twocolumn]{revtex4}
\usepackage{graphicx}
\usepackage{epstopdf}
\usepackage{dcolumn}
\usepackage{bm}
\usepackage{setspace}
\usepackage{psfrag}

\newcommand{\beq}{\begin{eqnarray}}
\newcommand{\eeq}{\end{eqnarray}}

\begin{document}
\title{The role of the ionic potential in high harmonic generation}

\author{D. Shafir$^1$, B. Fabre$^2$, J. Higuet$^2$, H. Soifer$^1$, M. Dagan$^1$, D. Descamps$^2$, E. M\'evel$^2$, S.
Petit$^2$, H. J. W\"orner$^3$, B. Pons$^2$, N. Dudovich$^1$ and Y. Mairesse$^2$
}

\affiliation{$^1$Department of Physics of Complex Systems, Weizmann
Institute of Science, Rehovot 76100, Israel\\
$^2$CELIA, UMR5107, Universit\'e de Bordeaux - CNRS - CEA, F33405 Talence, France\\
$^3$Laboratorium f\"ur Physikalische Chemie, ETH Z\"urich
Wolfgang-Pauli-Strasse 10, 8093 Z\"urich, Switzerland
}
\date{\today}

\begin{abstract}

Recollision processes provide a direct insight into the structure and dynamics of electronic wavefunctions. However, the strength of the process sets its basic limitations - the interaction couples numerous degrees of freedom. In this letter we decouple the basic steps of the process and resolve the role of the ionic potential which is at the heart of a broad range of strong field phenomena. Specifically we measure high harmonic generation from argon atoms. By manipulating the polarization of the laser field we resolve the vectorial properties of the interaction. Our study shows that the ionic core plays a significant role in all steps of the interaction. In particular, Coulomb focusing induces an angular deflection of the electrons before recombination. A complete spatio-spectral analysis reveals the influence of the potential on the spatio-temporal properties of the emitted light. 
\end{abstract}
\maketitle

Electron scattering is one of the main tools to probe the structure of matter. It has been successfully used to resolve the structure of solids \cite{solids}, molecular \cite{molecular} and biological \cite{biological} complexes. Recent advances in strong field light-matter interactions have opened a window into a new class of phenomena where electron scattering is initiated and manipulated by strong laser fields. In these experiments, the strong field leads to tunnel ionization which splits the electronic wavefunction into two parts: a bound state, associated with the hole left behind the
escaping electron, and a continuum state. The free part of the electron wavepacket is accelerated in the laser field and can recollide with the parent ion. Recollision can lead to the emission of optical radiation \cite{Cor93}. This process is known as high-order harmonic generation (HHG) \cite{brabec}.

In recent years, HHG has been the subject of numerous studies in which the harmonic spectrum was used to retrieve both the spatial \cite{Lein02,Ita04, Haes09, Levesque07,Shafir09} and temporal \cite{Bak06} characteristics of the electronic states participating in the process. When tunneling occurs from more than one orbital, a "hole" wavepacket is launched \cite{Smirnova09}. Furthermore, a significant population transfer induced by the strong laser field may evolve during the optical cycle \cite{Mairesse10}. When the electron recollides it maps the evolved hole wavefunction into the harmonic spectrum. These phenomena are at the heart of attosecond science. Their observation, so far, has been rather limited albeit they are likely present in various complex systems such as large molecules and clusters.

The main challenge is related to the complexity of the interaction. All degrees of freedom are coupled by the strong field light-matter interaction. How can we isolate the required information? One approach is to oversimplify the description. A commonly applied example is the Strong Field Approximation (SFA) \cite{Lewen94}, which neglects the influence of the ionic potential not only on the free electron trajectory but also on the structure of the recolliding wavepacket by treating it as a plane wave (plane wave approximation, PWA). In this letter we propose a different approach for probing recollision processes, demonstrating the ability to isolate the role of the ionic potential. The ionic potential is in fact involved in all stages of HHG. It dictates the structure of the involved bound state, defines the tunneling process \cite{Pfeiffer}, and modifies both the structure and dynamics of the recolliding electron. Isolating its role is a nontrivial challenge since the harmonic emission encodes the spatial and temporal properties of all the three steps involved in the generating process. A direct and unambiguous probing thus relies on the following requirements: 1. The study must be performed with systems in which the bound state remains static between ionization and recombination. 2. We have to control the recolliding wavefunction separately from the bound state. 3. We need to resolve the \textit{vectorial} properties of the induced dipole moment, modified by the ionic potential, eliminating the contribution of the tunneling probability and the recollision efficiency.

We address all the above requirements. We perform our study with argon atoms, whose ground state can be represented by field-free p orbitals even in the presence of a strong laser field \cite{Farrel11}. Second, we control the continuum part of the electronic wavefunction by inducing the interaction with an elliptically polarized field. Finally, we probe the vectorial properties of the interaction by measuring the polarization vector of the harmonics. For each harmonic, we separate contributions from two (long and short) delayed electron trajectories, which accordingly, allow us to scan not only the structure but also the dynamics of the system. Such separation is achieved via a complete spatio-spectral analysis. A semi-classical analysis of our measurements reveals the effect of the ionic potential on the structure of the recolliding electron wavepacket. Furthermore, we show that the longest electron trajectories are subject to a dynamical effect from the ionic potential, referred to as Coulomb focusing \cite{Brabec96}.

\begin{figure}
\begin{center}
\includegraphics[width=7.5cm,clip]{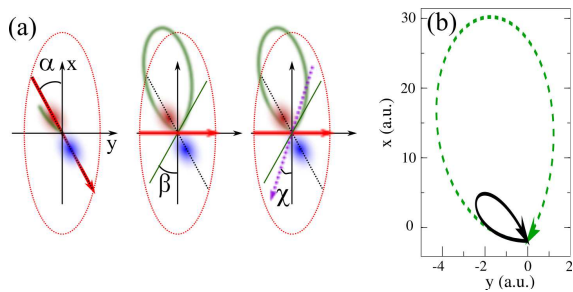}
\caption{(a) High harmonic generation in an elliptical field.
The red arrow represents the rotating electric field, the green line is the electron trajectory, and the purple arrow indicates the polarization direction of the high harmonics.
(b) Calculated classical short (continuous) and long (dashed)
electron trajectories leading to emission of harmonic 17 in argon
at 1.2$\times 10^{14}$ W.cm$^{-2}$ and $\epsilon_p=0.2$, in the
Strong Field Approximation.} \label{fig: pic}
\end{center}
\end{figure}

In many high harmonic spectroscopy experiments, the angle between the bound state and the recolliding electron wavepacket is defined by aligning the considered molecules. In order to probe the molecules from different angles, it is possible to change this alignment angle. However, this is not equivalent to simply changing the recollision direction with respect to a fixed bound state: rotating the molecule can modify the structure of the tunneling electron wavepacket, the tunneling probability or the bound state from which the electrons preferentially tunnel \cite{Smirnova09}.
This problem can be overcome by decoupling the recolliding wavefunction from the bound state through the control of its interaction with the laser field. Such a scheme has been recently demonstrated using a two color orthogonally polarized field \cite{Shafir09}. Here we use an alternative approach, producing high harmonics with elliptically polarized light (Fig. \ref{fig: pic}). Assuming an adiabatic response, tunnel ionization occurs along the
instantaneous polarization of the driving laser field. A quantization axis aligned in the direction $\alpha$ of the electric field at the time of ionization is then selected \cite{Young06, Loh07}. The elliptically polarized field shifts the freed electron in the lateral direction. Due to the
initial transverse momentum distribution, part of the electron wavepacket can recollide with its parent ion. The recollision occurs at an angle $\beta$ with respect to the main axis of the laser polarization ellipse $x$. Recollision leads to the emission of harmonic radiation polarized along the angle $\chi$. For spherically symmetric bound states the polarization angle is aligned along the recollision angle $\beta=\chi$. If the bound state is not spherically symmetric then the polarization angle $\chi$ is rotated from $\beta$ according to the structure of the continuum state and the bound state \cite{Shafir09}. By controlling the ellipticity of the IR field we control the relative angle between ionization and recollision.

The experiments were performed using the 1kHz 800 nm 35 fs Aurore laser system from CELIA. The ellipticity of the generating field is controlled by rotating a zero-order half-wave plate in front of a zero-order quarter-wave plate. The laser pulses are focused into a 1 kHz pulsed argon gas jet. We analyze the produced harmonics with an XUV flat field spectrometer. We optimize the laser focus position with respect to the gas jet to ensure proper
spectral and spatial separation of short and long trajectories on the detector due to phase matching conditions \cite{Heyl}. The polarization state of the high order harmonics is analyzed by inserting a 45$^\circ$ incidence unprotected silver mirror between the grating and the detector. The mirror reflectivity is $\sim30$ times higher for
S-polarized radiation than for P. By rotating the relative angle between the polarizer and the generating field's main axis, we observe a cosine squared modulation of the harmonic signal (Malus' law) from which we extract the polarization angle of each harmonic \cite{Ant96, MairesseNJP}. 

\begin{figure}[h]
\begin{center}
\includegraphics[width=0.4\textwidth]{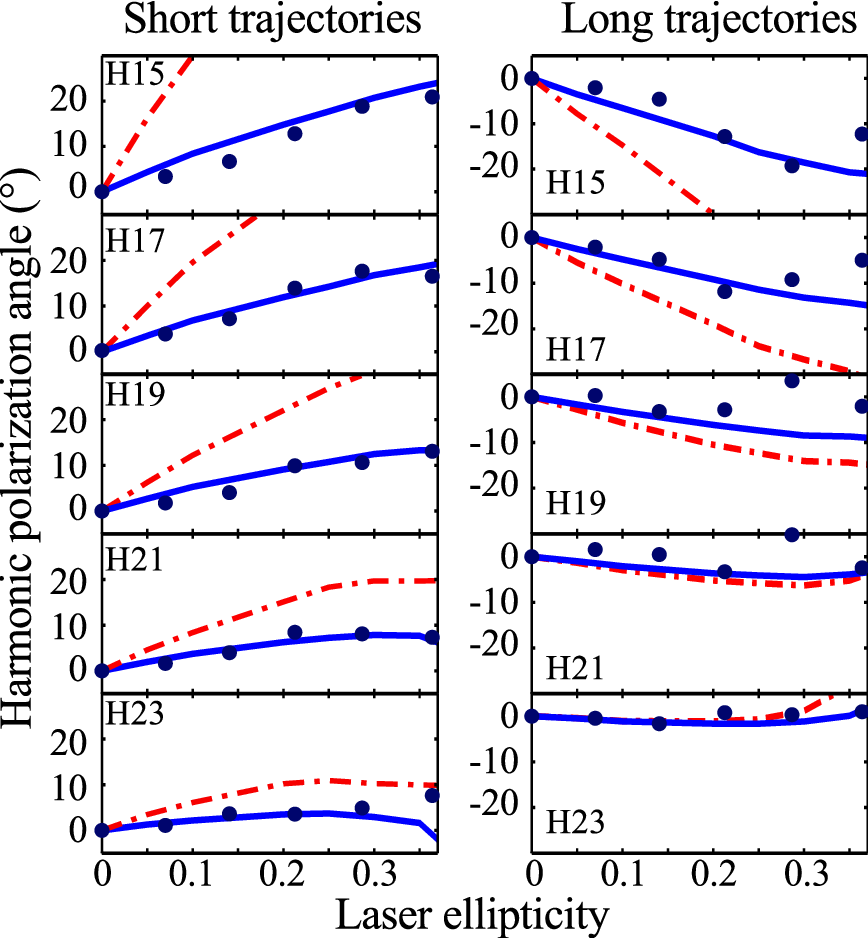}
\caption{High harmonic polarization angles in argon at 1.2$\times 10^{14}$ W.cm$^{-2}$, for short (left) and long (right) trajectories. Experimental results (black dots) and theoretical
CTMC-QUEST calculations using scattering continuum states (solid blue lines) and plane waves (dot-dashed red lines).}\label{results}
\end{center}
\end{figure}

For each harmonic, the signal from short and long trajectories is extracted by averaging the polarization angle over two separated areas on the detector. The resulting polarization angles $\chi$ obtained at $I\approx1.2\times 10^{14}$ W/cm$^2$ are shown in Fig. \ref{results} as a function of the laser ellipticity $\epsilon_p$. Several general features can be observed: (i) the polarization angle $\chi$ generally increases with $\epsilon_p$, (ii) the polarization angles measured from short and long trajectories have opposite signs (iii) for the highest harmonic orders $\chi$ remains close to zero for all ellipticities.

All these features can be qualitatively understood from classical analysis. Observation (i) can be easily explained: increasing the ellipticity increases the lateral displacement leading to larger recollision angles and thus to larger $\chi$. Feature (ii) can be explained considering the classical electron trajectories presented in Fig. \ref{fig: pic}(b). To have a closed trajectory, the electron must be launched into the continuum with an initial transverse momentum ($p_{y}^{ini}<0$) counteracting the accumulated momentum ($p_{y}^{field}>0$) . The final transversal momentum is dictated by the two components according to: $p_{y}^{tot}=p_{y}^{ini}+p_{y}^{field}$. Long trajectories, which interact with almost a complete cycle of the laser field, have a vanishing $p_{y}^{field}$, therefore $p_{y}^{tot}\approx p_y^{ini}<0$. By contrast, in short trajectories $p_{y}^{field}$ is significantly larger than $p_{y}^{ini}$, leading to $p_{y}^{tot}>0$. Opposite transverse momenta lead to opposite polarization angles for long and short trajectories as observed in (ii) (see also \cite{Strelkov}). High harmonics close to the cut-off are emitted by coalescing short and long trajectories for which $p_{y}^{ini}$ is almost totally counterbalanced by $p_{y}^{field}$. Accordingly, these trajectories recombine with $\beta \sim 0$ and feature (iii) follows.

Measuring $\chi$ provides a direct insight into the vectorial properties of the interaction. The value of $\chi$ relies on the relative orthogonal recombination dipole components and is intrinsically insensitive to the tunneling probability or the recollision efficiency. Therefore, it allows us to isolate the effect of the ionic potential, which we analyze hereinafter using a semi-classical theoretical approach.

Our theoretical model, CTMC-QUEST \cite{Higuet11}, combines a statistical description of electron paths in a Classical Trajectory Monte Carlo (CTMC) framework and a quantum mechanical description of the recombination process using QUantum  Electron Scattering Theory (QUEST) \cite{Le09}. In the framework of the single active electron approximation, CTMC employs an $N$-point discrete representation of the electron phase space distribution, $\varrho({\bf r},{\bf p},t)=\frac{1}{N} \times \sum_{j=1}^N{\delta({\bf r}-{\bf r}_j(t)) \delta({\bf p}-{\bf p}_j(t))}$, in terms of $N$ noninteracting  trajectories $\{ {\bf r}_j(t),{\bf p}_j(t) \}_{j=1,..,N}$. Electron dynamics are then monitored by the Liouville equation $\partial \varrho/\partial t=-[\varrho,\mathcal{H}]$ where $[\ ]$ is the Poisson bracket and $\mathcal{H}$ the effective one-electron Hamiltonian which
includes the model electron-ionic core interaction $V_c$ proposed in \cite{Muller} and the laser-atom interaction ${\bf r\cdot E}$ where ${\bf E}=(E_{x},E_{y})=E_0(\sin(\omega t), \epsilon_p\cos(\omega t))/\sqrt{1+\epsilon_p^2}$. The initial conditions are set by defining the distribution $\rho (r,p,t=0)$ as detailed in \cite{Higuet11}, to obtain spatial and momentum electron densities in close agreement with their quantum analog. The classical energy spread of the ground state allows mimicking tunneling in terms of above barrier transitions \cite{Solovev}. This correspondence has been clearly illustrated in stationary barrier cases \cite{ctmcstat} and quantitatively corroborated in non-stationary ones \cite{ctmcnonstat}.

Among the $N$ trajectories, we only select those which return at time $t=t_{rec}$ to the core with the angle $\beta$. Assuming pure selectivity of the bound state achieved by the ionization process, as schematically described in Fig. 1(a), these trajectories contribute to the dipole moment according to: 
\begin{equation}
d_{r_i}=\cos(\alpha)<\Psi_{{\bf k}}|r_i|\Psi_{3p_{x}}>
+\sin(\alpha)<\Psi_{{\bf k}}|r_i|\Psi_{3p_{y}}>,
\end{equation}\label{Eq. 1}
where $r_i=\{x,y\}$, $\Psi_{3p_{r_i}}$ are ground-state argon wavefunctions defined in the laboratory frame, and $\Psi_{{\bf k}}$ is the continuum wavefunction which describes the returning electron with momentum ${\bf k}$ \cite{Higuet11}. The angle of ionization, $\alpha=\tan^{-1}(E_{y}(t_{ion})/E_{x}(t_{ion}))$, determines the alignment of the quantization axis with respect to the $x$-axis of the laboratory frame (Fig. 1(a)). Drawing from Eq. (1), the HHG polarization angle is finally defined as $\chi=0.5 \tan^{-1}(|d_{x}||d_{y}|\cos\Phi/[|d_{x}|^2-|d_{y}|^2])$, where $\Phi$ is the phase difference between $d_{x}$ and $d_{y}$.

The value of ${\bf k}$ entering Eq. (1) is directly extracted from the returning CTMC trajectory. Evaluating $\alpha$ in the classical framework is more critical since $\alpha$ is determined by tunnel ionization. Even though it is possible to extract approximate ionization times from the classical trajectories \cite{note,Botheron09,Soifer,Botheron10}, we rely in practice on the SFA \cite{Ant96} to estimate the $t_{ion}$ values.


Figure 3 shows the polarization angles $\chi$ calculated by CTMC-QUEST, using continuum scattering states \cite{Higuet11}. The agreement with the measurements is very good. Based on this we have taken advantage of our semi-classical approach to investigate {\em separately} the static and dynamical
actions of the atomic potential $V_c$ on the harmonic polarization state.

First, we investigated the importance of the ionic potential on the structure of the recolliding electron wavefunctions. For that purpose we replaced the scattering waves $\Psi_{{\bf k}}$ by plane waves. The results are shown in Fig. 2 (dashed-dotted lines). These PWA calculation dramatically
fails in reproducing the observations.

\begin{figure}[h]
\begin{center}
\includegraphics[width=7.5cm,clip]{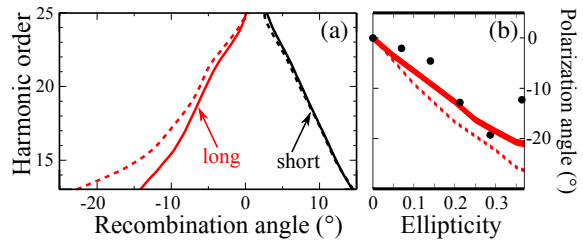}
\caption{(a) Electron recollision angles at $I=1.2\times 10^{14}$ W/cm$^2$ and $\epsilon_p=0.2$, using full (plain lines) and distorted (dashed lines) CTMC calculations for short and long trajectories. The distorted calculations neglect Coulomb focusing. (b): Polarization angles for long trajectories of harmonic 15: experiment (dots), full (plain line) and distorted (dashed line) CTMC-QUEST calculations.} \label{CTMC}
\end{center}
\end{figure}

Second, we studied the dynamical effect of the ionic potential on the electron trajectories in the continuum, by setting $V_c=0$ as soon as the trajectories turnaround and begin their return to the core. In Fig. 3 we compare the resulting recollision angles $\beta$ to those obtained with the full CTMC calculation. We observe that short trajectories are quite insensitive to the potential. By contrast, $V_c$ induces a significant decrease of the recombination angles associated with long trajectories. As we go to lower harmonic orders, the decrease in $\beta$ becomes more significant (up to $40 \%$ for the 13$^{th}$ harmonic). This angular shift is a signature of Coulomb focusing \cite{Brabec96}, an effect which has been suggested as an explaination of enhanced double ionization. To our knowledge, our results are the first demonstration of the effect of core focusing on HHG.

The different importance of Coulomb focusing to short and long trajectories can be intuitively explained. Short trajectories return to the ionic
core under the combined influence of the laser and ionic fields, which both accelerate them towards ${\bf r}={\bf 0}$. In contrast, long trajectories are slowed down since the two fields counteract each other before recombination \cite{Soifer,Botheron10,Gaarde}. Since cutting off a counteracting action has more
consequences than canceling a concurrent one, $V_c$ has more influence on the returning dynamics of the long trajectories than of the short ones.

Whereas Coulomb focusing has no effect on the polarization angle for short trajectories, it clearly reduces the polarization angle associated with the longest electron trajectories (Fig. \ref{CTMC}(b)). From a constructive point of view, this shows that the SFA can be safely used to describe above-threshold HHG associated with short trajectories, provided that accurate scattering states are used in the recombination step to overcome the above-mentioned PWA shortcomings. On the other hand, care must be taken when employing SFA to long trajectories even though accurate continuum states are used.

\begin{figure}
\begin{center}
\includegraphics[width=0.3\textwidth]{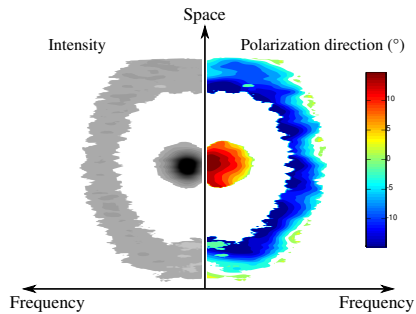}
\caption{ Spatio-spectral distributions of intensity and polarization angle in the far-field, for harmonic 15 driven by $\epsilon_p = 0.22$ . Short trajectories are collimated and long trajectories appear as a ring.} \label{fig: spatial}
\end{center}
\end{figure}

As in most high harmonic spectroscopy experiments, we have so far approximated the harmonic signal resulting from spatial and temporal averaging by the single atom response at a given effective laser intensity. The excellent agreement between our simulations and the experiment confirms the validity of this approach. However, we can take one step further by analyzing the entire spatio-spectral distribution of the polarization angle in our experiment. Figure \ref{fig: spatial} shows this distribution for harmonic 15 as it appears on our detector. A clear separation between short (central part) and long (ring) trajectories is observed \cite{Heyl}. Measuring the polarization state of each point on our detector, we resolve the complete spatio-spectral distribution. This analysis confirms the dramatic difference in polarization states of short and long trajectories. In addition, it provides a direct visualization of the intensity-dependence of the polarization angle. In our experiment the field intensity changes both temporally and spatially. As a result, the generation of each harmonic order is accompanied by a spatio-temporal chirp. The temporal chirp provides a simple mapping between time and frequency: the spectral maximum corresponds to the maximal temporal intensity, whereas the spectral wings correspond to the leading and falling edges of the pulse. A similar mapping occurs between the near field and far field's spatial profiles. Our measurements show that for each trajectory, the polarization angle decreases as we move away from the center of the spatio-spectral distribution, i.e. as the laser intensity decreases. This is consistent with the evolution of the recombination angle as we go from the plateau to the cutoff region (Fig. \ref{results}). Such an analysis can be easily applied in a broad range of experiments, resolving the spatio-spectral response of the interaction to a variable parameter --  pump-probe delay,  two-color phase or carrier-envelop phase.

The simultaneous spectral and spatial resolution of the detection allows us to measure both short and long trajectories, which can probe a system at a given electron energy from two different recollision angles, associated with two different recollision times. Such a measurement naturally enables the decoupling between structural and laser-induced information in high harmonic spectra. In this respect, our study contributes to HHG-based spectroscopic methods that aim at resolving the electronic structure and dynamics of atomic and molecular systems. 

Although we have focused on argon to reveal the role of a static ionic core, this work can straightforwardly be extended to more complex systems. One important aspect will then be the sensitivity to ionization times through the direction of the quantization axis $\alpha$, which should enable studying the influence of core polarization and Stark shifts on tunneling \cite{Pfeiffer}. This scheme can also be extended to resolve the influence of the ionic potential in molecular systems. Rotating the molecular axis will provide an additional information about the 2D properties of the potential. 

We acknowledge financial support from the ANR (ANR-08-JCJC-0029 HarMoDyn), the Conseil Regional d’Aquitaine (20091304003 ATTOMOL and COLA 2 n$^o$ 2.1.3-09010502), the European Union (228334 JRA-ALADIN Laserlab Europe II), the Minerva foundation, the Israeli Science foundation and the Crown Center of Photonics.


\begin{thebibliography}{}
\bibitem{note} The ionization time $t_{ion}$ associated to harmonic $q$ is determined in the CTMC framework as follows: among the trajectories freed from the target, we concentrate on those that return to the core with energy $E_i\pm \delta E_i$ compatible with $q$ ( $\delta E_i$ is typically 1/5 of the energy gap between successive harmonic orders). We consider that a trajectory is ionized, in a tunnel-like picture, as soon as it leaves the space spanned by the fundamental target state (for Ar, this bound space corresponds to a sphere with $r \sim 5$ a.u.). $t_{ion}$ is then defined as the mean value of the times when the trajectories belonging to a definite HHG wavepacket leave the sphere. As we previously found in \cite{Botheron09,Soifer,Botheron10} that the CTMC values of $t_{rec}$ are very close to those obtained by means of the SFA, we have presently found that the CTMC values of $t_{ion}$ are close to their SFA counterparts. However, since the CTMC determination of $t_{ion}$ requires a (cumbersome) analysis of all the wavepackets leading to HHG, we rely on the SFA to quickly estimate the $t_{ion}$ values.
\end{thebibliography}

\end{document}